\documentclass[aps,prb,twocolumn,superscriptaddress,showpacs,preprintnumbers]{revtex4}

\usepackage{graphicx}
\usepackage{dcolumn}
\usepackage{bm}

\begin{document}

\preprint{APS/123-QED}

\title{$^{31}$P NMR investigations on the ferromagnetic quantum critical system YbNi$_{4}$P$_{2}$}
\author{R. Sarkar}
\altaffiliation[]{rajib.sarkar@cpfs.mpg.de, rajibsarkarsinp@gmail.com}
\author{P. Khuntia}
\author{C. Krellner}
\author{C. Geibel}
\author{F. Steglich}
\author{M. Baenitz}
\affiliation{
Max-Planck Institute for Chemical Physics of Solids, 01187
Dresden, Germany}


\date{\today}

\begin{abstract}
We studied the new heavy-fermion system YbNi$_{4}$P$_{2}$, which
presents strong ferromagnetic correlations, using the local
$^{31}$P NMR probe over a wide field (0.2-8.6 \textsf{T}) and
temperature (1.8-200 K) range. $^{31}$P NMR Knight shift provides
the static spin susceptibility which tracks the bulk
susceptibility whereas the spin-lattice relaxation rate
$^{31}(1/T_{1}$) provide information about the fluctuations of
the Yb 4f moment. The Korringa law is valid over a wide range in
temperature and field. The Korringa product
$^{31}(1/T_{1}TK^{2}$S$_{0}$) $\ll$ 1 gives evidence for the
presence of strong ferromagnetic correlations. Over two decades
in temperature a $^{31}(1/T_{1}T) \sim T^{-3/4}$ behaviour was
found.
\end{abstract}

\maketitle
\begin{figure}
\includegraphics[scale=0.995]{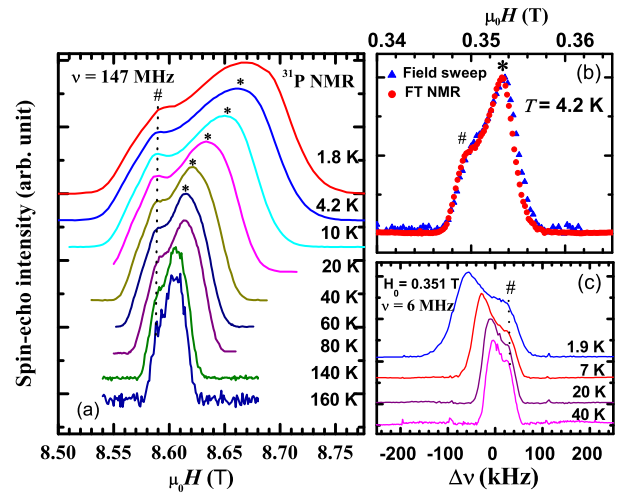}
\caption{\label{fig:spectra} (a) $^{31}$P field-sweep-NMR spectra
at 147 MHz at different temperatures (* marks the maximum
position used for the shift calculation whereas $\#$ marks the non
magnetic impurity Ni$_{3}$P ). (b) Comparison of the $^{31}$P
field sweep NMR  and FT NMR spectra taken at 4.2 K and 6 MHz. (c)
$^{31}$P FT NMR spectra at different temperatures at 6MHz
corresponding to the field 0.351 \textsf{T}.}
\end{figure}
\begin{figure}
\includegraphics[scale=1.0]{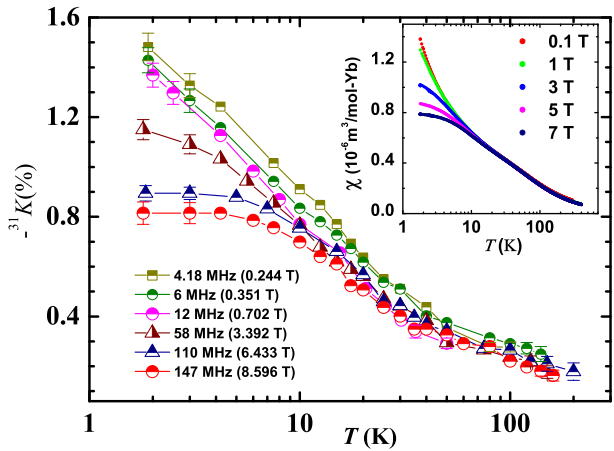}
\caption{\label{fig:ShiftvsT} $^{31}K(\%)$ as a function of
temperature for YbNi$_{4}$P$_{2}$ at different frequencies
(fields) as indicated. The field values are calculated from NMR
resonance frequency using ${^{31}\mathrm{\gamma}}/{2\pi}=$17.10
MHz/\textsf{T}. Inset shows the temperature vs. susceptibility
plot at different fields as indicated. The fields are chosen
similar to the NMR fields. }
\end{figure}

Recently quantum criticality (QC) has emerged as a central topic
especially in solid state physics.
While in the 4f and 5f based systems close to an antiferromagnetic
ordering QC is well established from experimental and theoretical
point of view, the observation of ferromagnetic quantum
criticality (FMQC) remains scarce and is mostly limited to 3d and
5f electron
systems.\cite{{Stewart-RMP-2001},{Nicklas-PRL-1999},{Sereni-PRB-2007},{Huy-PRB-2007},{Hilbert-rmdp-79-2007}}
Ferromagnetic quantum criticality (FMQC) has been discussed among
some 3d based weak itinerant ferromagnets like ZrZn$_{2}$
\cite{Hilbert-rmdp-79-2007} and NbFe$_{2}$
\cite{Brando-PRL-101-2008}, and 5f based systems like
UGe$_{2}$\cite{Taufour-PRL-105-2010} or UCoGe.
\cite{Ihara-PRL-105} In contrast among 4f systems it is rarely
discussed.\cite{{Sullow-PRL-82-1999},{Kirkpatrick-PRB-67-2003},{Yamamoto-PNAS-107-2010}}
\\
\indent In proximity to a quantum critical point (QCP),
unconventional power law behavior in the resistivity (
$\rho$(T)$\sim$ $T^{n}$ (n$<2$)), magnetic susceptibility
($\chi$$\sim$-$lnT$ or $T^{-n}$(n$<$1)) and specific heat
($C$$\sim$$lnT$ or $T^{-n}$(n$<$1)) could be observed
experimentally indicative of the deviations from standard
Fermi-liquid (FL) theory \cite{Baumbach-PRL-2010} and leads to the
concept of the non-Fermi-liquid (NFL) state. NFL behavior is fully
developed in the proximity of the QCP, but even far away from the
QCP the microscopic and macroscopic properties are influenced in
some temperature window. Therefore the normal state properties
must also be scrutinized in order to understand the diverse
properties of QC, especially with some microscopic tool.
Furthermore the lack of systematic NMR investigations on FMQC
systems in general to comprehend the spin dynamics also lead us
to investigate these type of systems in great detail.
\\
\indent The standard theory of Moriya for itinerant magnets
predicts for a 3D FM criticality close to the QCP for the spin
susceptibility (and spin-lattice relaxation rate)
$\chi_{\mathrm{Q}}\sim 1/T_{1}T \sim T^{-n}$ (n=4/3).\cite{{T.
Moriya-JPSJ-64-1995},{Moriya-JPSJ-1996},{T. Misawa-JPSJ-2009}}
Nonetheless such "clean" behaviour is very rare (see for example
UCoGe).\cite{Ihara-PRL-105} For more localized 4f based Ce or Yb
systems FMQC has been only poorly discussed and until now there is
no clear experimental evidence. Furthermore here, the formation
of a new, so called, "Quantum Griffiths " phase or a Kondo cluster
state originating from small disorder is discussed
.\cite{Sara-PRL-104-2010}  The ferromagnet CePd$_{1-x}$Rh$_{x}$
seems to be the prototype of this new sort of
magnetism.\cite{Westerkamp-PRL-102-2009} Here scaling in $C$,
$\chi$, $M$ and $1/T_{1}$ could be found which eventually should
lead to n$<$0.5 in 1/$T_{1}T$. Another possibility is the fragile
interplay of both, FM and AFM correlations in these systems. One
example for that is YbRh$_{2}$Si$_{2}$ where AF order at
$T_{\mathrm{N}}$ = 70 mK was found. Nonetheless NMR and ESR study
reveal the presence of additional FM correlations which are
promoted by magnetic fields. Here the system develops strong
ferromagnetic correlations evidenced by the NMR investigations
with a $^{29}(1/T_{1}T)$ $\sim T^{-0.5}$ power law associate to
the NFL behaviour.\cite{Ishida-PRL-2002} Despite the presence of
both ferro and antiferromagnetic correlations the system behaves
very local and the Korringa law is valid.
\\
\indent YbNi$_{4}$P$_{2}$ is a new heavy-fermion Kondo lattice,
discovered recently, with an extremely reduced Curie temperature
($T_{\mathrm{C}}$=0.17 K) due to strong Kondo screening
$T_{\mathrm{K}} \sim 8$ K in the close vicinity of a FM
QCP.\cite{Krellener-NJP-2011} The crystal structure is
quasi-one-dimensional with Yb$^{3+}$ chains along the c-axis of
the tetragonal unit cell. Between 50 and 300 K, the magnetic
susceptibility follows a Curie-Weiss law with stable Yb$^{3+}$
moments. A pronounced drop in the resistivity below 30 K
indicates the onset of coherent Kondo scattering, confirmed in a
pronounced minimum of the thermopower. Detailed low temperature
ac susceptibility measurements reveal a sharp FM transition at
$T_{\mathrm{C}}$ confirmed in the specific heat data which
presents a distinct $\lambda$ -type anomaly at $T_{\mathrm{C}}$.
Below $T_{\mathrm{C}}$, a heavy FL ground state is reflected in a
constant Sommerfeld-coefficient, $\gamma_0$=2
J/mol-K$^{2}$.\cite{Krellener-NJP-2011} Therefore this is a
promising candidate for a prototype, first reported Yb based FM
system close to quantum criticality.
\\
\indent In this communication we report $^{31}$P-NMR measurements
on the stoichiometric compound YbNi$_{4}$P$_{2}$. The Knight
shift, $^{31}K$, and the nuclear spin-lattice relaxation rate,
$^{31}(1/T_{1})$, were measured over a wide field range of 0.2-8.6
\textsf{T} to inspect the strong FM correlations suggested by the
bulk measurements. Being a local probe NMR can shed light on
microscopic magnetic properties by analyzing $^{31}K$ and
$^{31}(1/T_{1})$. $^{31}K$ gives information about the uniform
static spin susceptibility $\chi'$($\mathbf{q}$ = 0), while
$^{31}(1/T_{1}T)$ reveals the spin-fluctuation character from the
$\mathbf{q}$ averaged dynamical spin susceptibility
$\chi''$($\mathbf{q}$, $\mathrm{\omega}$). In the conventional FL
state, both $^{31}K$ and $^{31}(1/T_{1}T)$ are $T$ independent,
and the Korringa relation, $1/T_{1}TK^{2}$ =$S$= constant, is
valid. In the concept of renormalized heavy quasi particles the
ground state for $T\rightarrow 0$, far below the Kondo
temperature ($T_{\mathrm{K}}$), is the FL state where $K$ and
$1/T_{1}T$ are constant ("Kondo saturation") and the Korringa law
is valid too. Far above $T_{\mathrm{K}}$, $^{31}K$ and
$^{31}(1/T_{1}T)$ become $T$ dependent but if the coupling
mechanism between NMR nuclei and the local magnetic moment of the
4f ion is same for static and dynamic NMR responses, the Korringa
law could still be valid. Deviations from the $T$ independent
behaviour of $^{31}K$ and $^{31}(1/T_{1}T)$ at lower temperature
usually point towards the vicinity of a quantum critical point
and are interpreted as NFL behaviour, in analogy to the
unconventional power laws observed in bulk properties in the NFL
systems.
\\
\indent Figure~\ref{fig:spectra}(a) shows the $^{31}$P NMR powder
spectra taken at 147 MHz. The powder spectra is a superposition of
two lines. One line (marked by $\#$) shows no shift with
temperature and it is associated to a small amount of nonmagnetic
impurity phase ($<4\%$) Ni$_{3}$P. The main line comes from
$^{31}$P in YbNi$_{4}$P$_{2}$ and shows a magnetic negative shift
and a line broadening towards lower temperatures. A negative
shift is expected from the simple conduction electron
polarization model for Yb 4f ions.\cite{Carter-Bennett-1977} Due
to the fact that sizeable single crystals for NMR are not
available, the shift has been determined from powder results by
using the center position of the higher intensity peak (*) with
respect to the reference line marked by ($\#$). Making use of the
presence of non magnetic Ni$_{3}$P with $^{31}K$=0 gives very
accurate shift values specially for small fields where remanent
fields of the magnet create usually great problems in exact shift
determination. Furthermore we used H$_{3}$PO$_{3}$ with
$^{31}K$=0 as a reference compound for the absolute shift
determination at high fields. The aim of the paper was to probe
the critical fluctuation and/or the Kondo fluctuation in the zero
field limit. Therefore field sweep (FS) NMR measurements are
performed at very low fields. Here the line width is strongly
reduced and the FS method is at its limits. To overcome this
problem we switched to the more sensitive Fourier transform (FT)
NMR method. In Fig.~\ref{fig:spectra}(b) the comparison of the
$^{31}$P NMR spectra taken at FS (6 MHz) and FT method (0.351
\textsf{T}) are plotted after normalization to the field axis.
This plot confirms that these two different methods ultimately
give the same results. For the low frequencies 4 and 6 MHz we
used therefore only the FT-NMR method. As an example FT spectra
at 6 MHz at a center field H$_{0}$=0.351 \textsf{T} are shown in
Fig.~\ref{fig:spectra}(c).
\\
\indent Figure~\ref{fig:ShiftvsT} displays the $^{31}K(T)$ vs. $T$
plot. $^{31}K(T)$ shows a nice agreement with the bulk magnetic
susceptibility. At low fields (0.244 \textsf{T}, 0.351 \textsf{T}
and 0.702 \textsf{T}), $^{31}K$ increases monotonously with
lowering the temperature down to 1.8 K without sign of any
saturation. However at higher fields (6.433 \textsf{T} and 8.596
\textsf{T}), $^{31}K(T)$ starts to saturate towards lower
temperature. Additionally the onset of the saturation is shifted
towards higher temperatures with increasing fields.
Figure~\ref{fig:ShiftvsT} (inset) shows the susceptibility vs.
$T$ plot at different fields as indicated. This plot is
consistent with the $^{31}K(T)$ vs. $T$ plot. Therefore this rules
out the presence of any magnetic (FM and/or AFM) impurity
contribution in $\chi(T)$. The hyperfine coupling constant
($A_{\mathrm{hf}}$) is estimated by plotting $^{31}K(T)$ with
respect to bulk susceptibility (not displayed here). The value of
 $A_{\mathrm{hf}}$ is 592 Oe/$\mu_{\mathrm{B}}$ which is close
to the value obtained for
YbRh$_{2}$Si$_{2}$.\cite{Ishida-PRL-2002} The saturation of
$^{31}K$ below 8 K for 6.433 \textsf{T} and 8.596 \textsf{T} can
be interpreted as the polarization effect of the external field on
the Yb$^{3+}$ localized moment. At around 80 K a shoulder is
observed in the $^{31}K(T)$ vs. $T$ plot, which is likely caused
by crystal electric field (CEF) excitations. A similar feature is
also observed in the susceptibility data.
\begin{figure}
\includegraphics[scale=0.95]{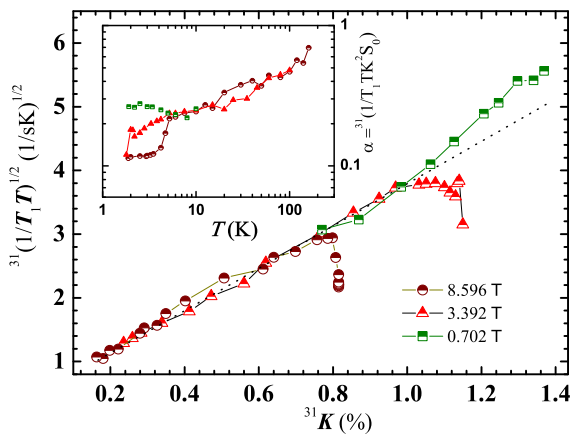}
\caption{\label{fig:T1TvsK} $^{31}$($1/\sqrt{T_{1}T}$) as a
function of $^{31}K(\%)$ at different fields as indicated. The
dotted line indicates the linear $^{31}K(T)$ dependence. At high
$^{31}$K values (low temperatures) the experimental data show an
upward turn for low fields due to critical fluctuations, and a
downward turn for high fields due to field polarised state. The
inset shows the $T$ dependence of the Korringa product
$\alpha=^{31}(1/T_{1}TK^{2}S_{0})$ (see text).}
\end{figure}
\\
\indent Now we present nuclear spin-lattice relaxation rate
$^{31}(1/T_{1})$ data on YbNi$_{4}$P$_{2}$.  $^{31}(1/T_{1})$
measurements were performed as a function of temperature at
different frequencies 12, 58 and 147 MHz (corresponding to the
fields 0.702 \textsf{T}, 3.392 \textsf{T} and 8.596 \textsf{T}
respectively) by exciting at the maximum of the anisotropic NMR
spectra (marked by $\star$ in Fig. 1). The nuclear magnetization
recovery curves could be fitted at any temperature and field with
a standard single exponential function expected for a I=1/2 NMR
nuclei. This indicates that the system has a single relaxation
channel for a particular field. Before starting the rather
complex discussion of the \textit{H}-and $T$-dependence of
$^{31}(1/T_{1})$ we first rise up the question whether this
system is an itinerant or a more localized system. For a
localized system the Korringa theory should be applicable,
whereas for an itinerant systems results should be discussed in
the framework of the Moriya theory. A first evidence for
YbNi$_{4}$P$_{2}$ being a local system comes from the
susceptibility data of Krellner \textit{et. al.} giving proof for
a full Yb$^{3+}$ moment above 150 K.\cite{Krellener-NJP-2011} The
ultimate NMR probe is the $^{31}K$ dependence of
$^{31}(1/T_{1}T)$. If $^{31}(1/T_{1}T)$ $\sim$ $^{31}K^{2}$ is
observed then Korringa law is valid, whereas a linear $^{31}K$
-dependence points towards an itinerant system where the Moriya
theory should be applied.\cite{T. Moriya-spinfluctuations-1985}

Fig. \ref{fig:T1TvsK} shows the $^{31}(1/\sqrt{(T_{1}T))}$ vs.
$^{31}K$ plot for three different fields. For high fields (8.596
\textsf{T} and 3.392 \textsf{T}) they follow almost linear
behavior, except at low temperature, where due to the field
polarized state, a bending occurs (see below). The dotted line in
Fig. \ref{fig:T1TvsK} is the guide to the linear dependency. The
almost linear relation between $^{31}(1/\sqrt{(T_{1}T))}$ and
$^{31}K$ indicate the validity of the Korringa law, which means
that one has to consider the localized moment framework. Even
though this was already evidenced by the bulk measurements, now
it is also clear from the viewpoint of a local picture. However,
at low fields (0.702 \textsf{T}) a clear upward deviation from the
linearity is observed, which is likely related to the development
of critical magnetic fluctuations originating from the fragile
interplay of Kondo and FM correlations.
\begin{figure}
\includegraphics[scale=1.05]{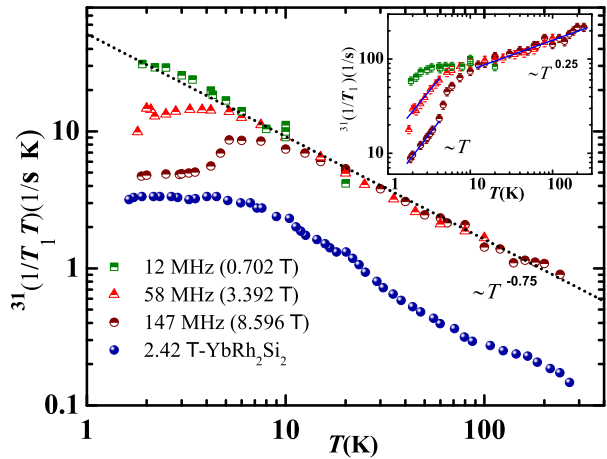}
\caption{\label{fig:T1TvsT} $^{31}$($1/TT_{1}$) vs. $T$ plot at
different fields as indicated. Blue sphere represent the
$^{29}$Si NMR data for YbRh$_{2}$Si$_{2}$ at 2.42 T (reference
\cite{Ishida-PRL-2002}) after multiplying by $(A_{\mathrm{hf}}^{2}
\gamma^{2})_{\mathrm{P}}/(A_{\mathrm{hf}}^{2}
\gamma^{2})_{\mathrm{Si}}$. Inset shows the $^{31}$($1/T_{1}$) as
a function of temperature at different fields as indicated.}
\end{figure}
It should be mentioned that the above described behaviour is
reminiscence of YbRh$_{2}$Si$_{2}$. There a similar behaviour is
observed but with the difference that AFM order shows up at low
$T$ ($T_{\mathrm{N}}$=70 mK). Therefore we have plotted in
Fig.~\ref{fig:T1TvsT} the temperature dependence of
$^{31}(1/T_{1}T)$ and the $^{31}(1/T_{1})$ on a double log scale
in the main panel and inset respectively together with $^{29}$Si
NMR data of YbRh$_{2}$Si$_{2}$ (at 2.42 \textsf{T}) taken from Ref
\cite{Ishida-PRL-2002}. Interestingly the results look very
similar to that of YbNi$_{4}$P$_{2}$ at around 3.392 \textsf{T}.
Considerable field (frequency) dependence of $^{31}(1/T_{1}T)$ is
observed below 10 K, in agreement with the $^{31}K$, $\chi(T)$
and $C(T)$ results. Above 10 K $^{31}(1/T_{1}T)$ follows a
$T^{n}$ power law with n smaller than -1 (n=$-\frac{3}{4})$ over
two decades in temperature. With further lowering the temperature
$^{31}(1/T_{1}T)$ deviate from this (n$=-\frac{3}{4}$) power law
and becomes constant leaving a broad maxima. The occurrence of
such a weak power law over a wide temperature and field range is
even though rare, nonetheless found for systems like USb$_{2}$,
CeCoIn$_{5}$, YbAuCu$_{4}$ and therefore could not be simply
justified as an accident.\cite{{YAMAMOTO-JPSJ-2007},
{Baek-PRB-2010},{Urbano-PRL-2007}} For local moment 4f system far
from the critical point such behaviour is unusual. For example in
a 4f heavy-fermion, like CeCu$_{2}$Si$_{2}$ (1/$T_{1}T$) levels
into a constant value ("Kondo saturation") just below the Kondo
temperature $T_{\mathrm{K}}$ and above $T_{\mathrm{K}}$ in most
cases a n = -1 power law (constant $1/T_{1}$ behaviour) is
observed.
\cite{Arts-CeCu2Si2-NMR-1983,Kawasaki-CeCu2Si2-NQR-2002} In
contrast to that in the two Yb based correlated system
YbRh$_{2}$Si$_{2}$ and YbNi$_{4}$P$_{2}$, there is no Kondo
saturation and a n $<$ 1 power law is observed. The absence of
the "Kondo saturation" might be related to the presence of
ferromagnetic correlations whereas the high temperature behaviour
might originate from the CEF splitting. The validity of the
Korringa law suggest that the non Curie Weiss behaviour of the
static susceptibility towards lower temperatures caused by CEF
splitting is responsible for the $1/T_{1}T$ behaviour.
\\
\indent For free electron metals the Korringa relation is given
by $1/T_{1}TK^{2}$ $=$ S$_{0}$ = $
\pi$$\hbar$$\gamma_{\mathrm{N}}^{2}
k_{\mathrm{B}}^{}$/$\mu_{\mathrm{B}}^{2}$ where
$\gamma_{\mathrm{N}}$ is the nuclear gyromagnetic ratio of
$^{31}$P and $k_{\mathrm{B}}$ is the Boltzman constant. Including
electronic correlations leads to the modified Korringa relation
$1/T_{1}TK^{2}$ $=S= \alpha \* S_{0}$. The so called Korringa
product $(1/T_{1}TK^{2}S_{0}) = \alpha $ is a very useful probe
for correlations, $\alpha=1$ indicates the absence of correlation,
whereas $\alpha
>1$ indicates AFM correlation and $\alpha <1$ FM correlation. For
the $^{31}$P nuclei $S_{0}$ is 0.623$\times10^{6}$ 1/sK while
experimentally we found $0.133\times10^{6}$ 1/sK at 2 K. This
gives a value of $\alpha$ (2 K) $\simeq$ 0.21 which indicates
ferromagnetic correlations like in YbRh$_{2}$Si$_{2}$ (
$\alpha$=0.11 at 100 mK) or like in CeFePO
($\alpha$=0.065).\cite{{Ishida-PRL-2002},{E. M.
Bruning-prl-2008}} This is also consistent with the strongly
enhanced Sommerfeld-Wilson ratio $W_{T\rightarrow ~0.3~\mathrm{K}}
\cong 20 $ found for YbNi$_{4}$P$_{2}$.\cite{Krellener-NJP-2011}
The inset of the Fig.~\ref{fig:T1TvsK} shows the temperature and
field dependency of $\alpha$.
\\
\indent In summary, we have presented a $^{31}$P NMR study on the
newly discovered heavy-fermion Kondo lattice system
YbNi$_{4}$P$_{2}$ to shed some light into its microscopic
properties. At low fields $^{31}K(T)$ and $^{31}(1/T_{1}T)$ are
showing no sign of saturation towards low temperatures, in
contrast to a heavy FL state well below $T_{\mathrm{K}}$=8 K. On
the contrary the Korringa law is valid over a wide field and
temperature range. Below 10 K and at low fields the breakdown of
the Korringa law points towards the onset of critical
ferromagnetic fluctuation. In contrast to Ce heavy-fermion
systems, but very similar to YbRh$_{2}$Si$_{2}$,
$^{31}(1/T_{1}T)$ shows a weak power law with $1/T_{1}T \sim
T^{-n}$ with n $<$ -1 down to the lowest temperatures. We
speculate that the $^{31}(1/T_{1}T)$ behaviour could be
associated with the CEF splitting changing the effective magnetic
moment of the system. Moreover, the value of the Korringa product
being smaller than one strongly suggests the presence of FM
correlations. At low fields, $^{31}(1/T_{1}T)$ results indicate
the development of critical fluctuations. Until now there is
still a lack of complete understanding of YbNi$_{4}$P$_{2}$. NMR
measurements should be extended towards lower temperature and
single crystals are required to investigate the magnetic
anisotropy. Furthermore, inelastic neutron-scattering studies are
required to investigate the \textbf{q}-dependence of the
fluctuations. Interestingly for YbRh$_{2}$Si$_{2}$ inelastic
neutron studies strongly reveal the presence of two relaxation
channels. Experimentally at low temperatures the quasielastic
linewidth has been found to have two components, one constant
component ( Kondo fluctuation) and one depending linear on $T$
(inter-site fluctuations).\cite{Stockert-371-2007} It would be
rather interesting to see if neutron studies on YbNi$_{4}$P$_{2}$
shows similar features.


\begin{thebibliography}{10}

\bibitem{Stewart-RMP-2001}
 G. R. Stewart Rev. Mod. Phys. \textbf{73}, 797 (2001).
\bibitem{Nicklas-PRL-1999}
 M. Nicklas, M. Brando, G. Knebel, F. Mayr, W. Trinkl, and A. Loidl, Phys. Rev. Lett. \textbf{82}, 4268 (1999).
\bibitem{Sereni-PRB-2007}
J. G. Sereni, T. Westerkamp, R. K\"{u}chler, N. Caroca-Canales,
P. Gegenwart, and C. Geibel, Phys. Rev. B \textbf{75}, 024432
(2007).
\bibitem{Huy-PRB-2007}
N. T. Huy, A. Gasparini, J. C. P. Klaasse, A. de Visser, S.
Sakarya, and  N. H. van Dijk, Phys. Rev. B \textbf{75}, 212405
(2007).
\bibitem{Hilbert-rmdp-79-2007}
H. v. L\"{o}hneysen, A. Rosch, M. Vojta, and Peter W\"{o}lfle
Rev.  Mod. Phys. \textbf{79}, 1015 (2007).
\bibitem{Brando-PRL-101-2008}
M. Brando, W. J. Duncan, D. Moroni-Klementowicz, C. Albrecht, D.
Gr\"{u}ner, R. Ballou, and F. M. Grosche, Phys. Rev. Lett.
\textbf{101}, 026401 (2008).
\bibitem{Taufour-PRL-105-2010}
 V. Taufour, D. Aoki, G. Knebel, and J. Flouquet,
Phys. Rev. Lett. \textbf{105}, 217201 (2010).
\bibitem{Ihara-PRL-105}
Y. Ihara, T. Hattori, K. Ishida, Y. Nakai, E. Osaki, K. Deguchi,
N. K. Sato, and I. Satoh, Phys. Rev. Lett. \textbf{105}, 206403
(2010).
\bibitem{Sullow-PRL-82-1999}
S. S\"ullow, M. C. Aronson,  B. D. Rainford, and P. Haen, Phys.
Rev. Lett. \textbf{82}, 2963-2966 (1999).
\bibitem{Kirkpatrick-PRB-67-2003}
T. R Kirkpatrick, and  D. Belitz,
 Phys. Rev. B \textbf{67}, 024419 (2003).
\bibitem{Yamamoto-PNAS-107-2010}
S. J. Yamamoto, and Q. Si, Proc. Nat. Acad. Sci. \textbf{107},
15704-15707 (2010).
\bibitem{Baumbach-PRL-2010}
R. E. Baumbach, J. J. Hamlin,  L. Shu,  D. A. Zocco,  J. R.
O'Brien,  P.-C. Ho, and M. B. Maple, Phys. Rev. Lett.
\textbf{105}, 106403 (2010).
\bibitem{T. Moriya-JPSJ-64-1995}
T. Moriya and T. Takimoto, J. Phys. Soc. Jpn. \textbf{64}, 960
(1995).
\bibitem{Moriya-JPSJ-1996}
A. Ishigaki and T.Moriya  J. Phys. Soc. Jpn \textbf{65}, 3402
(1996).
\bibitem{T. Misawa-JPSJ-2009}
T. Misawa, Y. Yamaji, and M. Imada, J. Phys. Soc. Jpn.
\textbf{78}, 084707 (2009).
\bibitem{Sara-PRL-104-2010}
S. Ubaid-Kassis, T. Vojta, and A. Schroeder, Phys. Rev. Lett.
\textbf{104}, 066402 (2010).
\bibitem{Westerkamp-PRL-102-2009}
T. Westerkamp, M. Deppe, R. K\"{u}chler, M. Brando, C. Geibel, P.
Gegenwart, A. P. Pikul, and F. Steglich, Phys. Rev. Lett.
\textbf{102}, 206404 (2009).
\bibitem{Ishida-PRL-2002}
K. Ishida,  K. Okamoto, Y. Kawasaki, Y. Kitaoka,  O. Trovarelli,
C. Geibel, and F. Steglich, Phys. Rev. Lett. \textbf{89}, 107202
(2002).
\bibitem{Krellener-NJP-2011}
C. Krellner, S. Lausberg, A. Steppke, M. Brando, L. Pedrero, H.
Pfau, S. Tenc\'{e}, H. Rosner, F. Steglich, and C. Geibel, New
Journal of Physics \textbf{13}, 103014(2011).
\bibitem{Carter-Bennett-1977}
G. C. Carter, L. H. Bennett, and D. J. Kahan, Metallic Shifts in
NMR I (Pergamon Press, Oxford, 1977).
\bibitem{YAMAMOTO-JPSJ-2007}
A. Yamamoto, S. Wada, and J. L. Sarrao, J. Phys. Soc. Jpn
\textbf{76}, 063709 (2007).
\bibitem{Baek-PRB-2010}
S.-H. Baek, N. J. Curro, H. Sakai, E. D. Bauer, J. C. Cooley, and
J. L. Smith, Phys. Rev. B  \textbf{81}, 054435 (2010).
\bibitem{Urbano-PRL-2007}
R. R. Urbano, B.-L. Young, N. J. Curro, J. D. Thompson, L. D.
Pham, and Z. Fisk, Phys. Rev. Lett. \textbf{99}, 146402 (2007).
\bibitem{T. Moriya-spinfluctuations-1985}
T. Moriya: Spin Fluctuations in Itinerant Electron Magnetism
(Springer, Berlin, 1995).

\bibitem{Kawasaki-CeCu2Si2-NQR-2002}
Y. Kawasaki, K. Ishida, K. Obinata, K. Tabuchi, K. Kashima, Y.
Kitaoka, O. Trovarelli, C. Geibel, and F. Steglich Phys. Rev. B
\textbf{66}, 224502 (2002).

\bibitem{Arts-CeCu2Si2-NMR-1983}
J. Aarts, F. deBoer, and D. E. MacLaughlin, Physica (Amsterdam)
\textbf{121B+C}, 162 (1983).


\bibitem{E. M. Bruning-prl-2008}
E. M. Br\"{u}ning, C. Krellner, M. Baenitz, A. Jesche, F.
Steglich, and C. Geibel, Phys. Rev. Lett. \textbf{101}, 117206
(2008).
\bibitem{Stockert-371-2007}
O. Stockert, M.M. Koza, J. Ferstl, C. Geibel, F. Steglich,
Science and Technology of Advanced Materials \textbf{8}, 371-375
(2007).


\end{thebibliography}

\end{document}